\documentclass[english,aps,prl,reprint,superscriptaddress]{revtex4-1}
\usepackage[T1]{fontenc}
\usepackage[latin9]{inputenc}
\usepackage{units}
\usepackage{amsmath}
\usepackage{amssymb}
\usepackage{graphicx}
\usepackage{wasysym}
\usepackage{esint}
\usepackage{babel}
\begin{document}
\title{Phase Transition and Diffusion in an Adiabatic Bouncer Model }
\author{Luiz Antonio Barreiro}
\email{luiz.a.barreiro@unesp.br}

\address{{\small{}Institute of Geosciences and Exact Sciences, }~\linebreak{}
{\small{}Physics Department, São Paulo State University (Unesp), CEP
13506-900, Rio Claro, São Paulo, Brazil}}
\begin{abstract}
This study examines anomalous diffusion and dynamical phase transitions
in a nonlinear bouncer model with short-range interactions leading
to velocity-dependent (adiabatic) collisions. By varying a control
parameter, transitions between superdiffusion, normal diffusion, subdiffusion,
and complete dynamical freezing are observed. The phase behavior is
characterized via scaling laws and critical exponents, providing a
robust framework for understanding the underlying dynamics.
\end{abstract}
\maketitle
\textit{Introduction} - Nonlinear dynamical systems display a wide
range of complex behaviors, including bifurcations, chaos, anomalous
diffusion, and phase transitions \citep{Strogatz2018,Ott2002}. Among
these, transport phenomena such as diffusion often deviate from classical
Brownian motion when governed by nonlinear or chaotic dynamics. In
such cases, deterministic chaos and nonlinearity can lead to anomalous
diffusion, where the mean square displacement scales as $\langle x^{2}(t)\rangle\sim t^{\mu}$
with $\mu\neq1$ \citep{Klages2008,Metzler2000}.

Dynamical phase transitions\textemdash abrupt changes in statistical
behavior driven by control parameters\textemdash are marked by shifts
in the scaling exponent $\mu$, indicating regimes of superdiffusion
$(\ensuremath{\mu>1})$, normal diffusion $(\ensuremath{\mu=1})$,
subdiffusion $(\ensuremath{\mu<1})$, or complete suppression of transport
(freezing) \citep{Venegeroles2009,Zaburdaev2015}.

Models such as bouncers \citep{Liebchen2011,Leonel2004}, kicked rotators
\citep{Chirikov1979}, and nonlinear maps \citep{Zaslavsky2002} have
been widely used to study how phase space structure influences diffusion
and critical behavior.

This work investigates a modified bouncer model incorporating an adiabatic
parameter that accounts for molecular-scale interactions during collisions.
The classical bouncer model\textemdash describing a particle colliding
elastically with an oscillating surface under gravity\textemdash serves
as a key system for studying Fermi acceleration and chaotic transport.
Introducing velocity-dependent restitution or finite-time interactions
extends the model\textquoteright s physical relevance, allowing for
the analysis of anomalous transport, transitions to a frozen regime,
and critical phenomena. Emphasis is placed on scaling laws and critical
exponents that characterize the system near the transition point. 

\textit{Collision process} - The analysis begins with the classical
bouncer model, which describes a particle subject to a uniform gravitational
field $g$, undergoing successive collisions with a platform that
oscillates harmonically in time. In the reference frame of the oscillating
surface, the velocity of the particle relative to the platform at
the instant of collision is expressed as $v_{i}'=v_{i}-V,$ where
$v_{i}$ is the particle's velocity just before the collision in the
lab frame, and $V$ is the instantaneous velocity of the surface.
Assuming elastic collisions $(v'_{f}=-v'_{i})$ and transforming back
to the laboratory frame, we obtain
\begin{equation}
v_{f}=\left|v_{i}\right|+2V.\label{velocity}
\end{equation}

To improve the physical realism of the bouncer model, molecular-scale
effects can be included by accounting for short-range repulsive forces,
such as dipole interactions \citep{Huang1987}. Unlike idealized instantaneous
collisions, the interaction occurs over a finite time $\tau\sim\ell/v_{n}$,
where $\ell$ is a characteristic interaction length and $v_{n}$
the particle's incident velocity. During this interval, the wall's
velocity is effectively time-averaged over the contact duration $\tau$,
yielding {\small{}
\begin{equation}
\langle V\rangle=\frac{1}{\tau}\intop_{t_{c}}^{t_{c}+\tau}\zeta\omega\cos(\omega t)dt\approx\zeta\omega\left[\cos(\omega t_{c})-\frac{\omega\ell}{2v_{i}}\sin(\omega t_{c})\right],\label{velFinitTime}
\end{equation}
}where $t_{c}$ is the collision time, and $\zeta$, $\omega$ characterize
the amplitude and frequency of the wall's motion. The first term represents
the instantaneous velocity of the wall, while the second accounts
for the time-dependent force during contact.

This finite-time correction becomes especially relevant for slow-moving
particles ($v_{i}\ll1$), introducing significant deviations from
elastic behavior. For fast particles ($v_{i}\gg1$), the interaction
becomes effectively instantaneous, recovering the adiabatic limit.

\textit{The Map} - Based on the Eqs (\ref{velocity}) and (\ref{velFinitTime}),
we construct a discrete map describing the velocity update after each
collision. At the $(n+1)$-th bounce, the particle velocity is given
by 
\begin{equation}
v_{n+1}=\left|\left[1+\frac{\zeta\sin(\omega t_{n+1})}{v_{n}^{z}}\right]v_{n}\right|,
\end{equation}
where the $z$ parameter controls the degree of adiabaticity. For
$z=1$, the interaction mimics instantaneous contact; for $z=2$,
it models a finite-duration force response. The modulus dictates that
a particle with negative velocity is absorbed by the surface, and
a new particle is injected with equal magnitude but opposite velocity.

Between collisions, the particle undergoes free fall in a uniform
gravitational field. The time between bounces is $t_{n+1}-t_{n}=\frac{2v_{n}}{g}.$
Introducing the dimensionless variables $\phi_{n}=\omega t_{n}$ and
$V_{n}=\omega v_{n}/g$, the dynamics is governed by the two-dimensional
map 
\begin{equation}
\begin{aligned}\phi_{n+1} & =\phi_{n}+2V_{n},\\
V_{n+1} & =\left|V_{n}+\frac{\bar{\zeta}}{V_{n}^{z-1}}\sin(\phi_{n+1})\right|,
\end{aligned}
\label{Smap-1-1}
\end{equation}
where an effective parameter $\bar{\zeta}=\zeta(\omega/g)^{z}$ was
defined. This map captures the transition from chaotic diffusion to
localized dynamics, depending on the choice of $z$ and $\bar{\zeta}$. 

\textit{Diffusion Process} - It is possible to use an analytical argument
to predict the unlimited diffusion of energy through the Fermi acceleration
mechanism \citep{Leonel2015}. Let us consider an ensemble of particles
with initially low velocities, which means an ensemble characterized
by a low temperature, but sufficiently high to neglect quantum effects.
Squaring the second equation in mapping (\ref{Smap-1-1}), and making
an ensemble average over $\phi\in[0,2\pi]$, yields
\begin{equation}
\left\langle V_{n+1}^{2}\right\rangle =\left\langle V_{n}^{2}\right\rangle +\frac{1}{2}\bar{\zeta}^{2}\left\langle V_{n}^{2-2z}\right\rangle .
\end{equation}
In the asymptotic limit of large $n$, the relationship can be expressed
as
\begin{eqnarray}
\frac{\partial\left\langle V_{n}^{2}\right\rangle }{\partial n} & \simeq & \frac{\left\langle V_{n+1}^{2}\right\rangle -\left\langle V_{n}^{2}\right\rangle }{(n+1)-n}=\frac{\bar{\zeta}^{2}}{2}\left\langle V_{n}^{2-2z}\right\rangle .
\end{eqnarray}

To advance further, let us assume a Gaussian form for the distribution,
with an anomalous diffusion $\Psi_{\mu}(V,n)=\sqrt{\frac{a}{\pi}}\frac{1}{n^{\mu}}e^{-a\left(\frac{V}{n^{\mu}}\right)^{2}}$
\citep{Cecconi2022,Boscolo2023}. Under this assumption, it is possible
to write $\left\langle V_{n}^{2-2z}\right\rangle =\intop_{-\infty}^{\infty}V^{2-2z}\Psi_{\mu}(V,n)\,dV=\frac{2^{1-z}}{\sqrt{\pi}}\varGamma\left(\frac{3}{2}-z\right)\left\langle V_{n}^{2}\right\rangle ^{1-z}$
leading to a differential equation in $\left\langle V_{n}^{2}\right\rangle $.
The solution to this equation can then be expressed as{\small{}
\begin{equation}
\left\langle V_{n}^{2}\right\rangle =\left\{ V_{0}^{2z}+z\varGamma\left(\frac{3}{2}-z\right)\frac{\bar{\zeta}^{2}}{\sqrt{\pi}}n\right\} ^{\nicefrac{1}{z}}\sim\left(\beta\,n\right)^{\nicefrac{1}{z}}.\label{V2-2}
\end{equation}
where we have defined} {\small{}$\beta(z,\bar{\zeta})=z\varGamma\left(\frac{3}{2}-z\right)\frac{\bar{\zeta}^{2}}{\sqrt{\pi}}$
}and $V_{0}$ is the initial velocity. 

The results presented in (\ref{V2-2}) demonstrate that the diffusion
behavior of the system is critically dependent on the value of the
adiabatic parameter. In this context, normal diffusion corresponds
to a mean square velocity (MSV) that grows linearly with the number
of iterations $n$ $(z=1)$. Superdiffusion is characterized by an
MSV that increases faster than linearly with $n$ $(z<1)$, typically
associated with long-range correlations or persistent motion. Conversely,
subdiffusion is identified by an MSV that grows more slowly than linearly,
often resulting from dynamical trapping or stability effects in phase
space $(z>1)$. Thus, the $z$ parameter functions as a critical control
parameter, governing the transition between superdiffusive, diffusive,
and subdiffusive regimes in the system's energy transport. This parameter
effectively modulates the rate and nature of energy transfer within
the system, influencing the overall dynamic behavior and allowing
for precise manipulation of diffusion processes.

\textit{Trapping Criterion at Low Velocity} - The derivation of these
diffusion regimes relies on two principal assumptions: The phase variable
$\phi$ is uncorrelated and uniformly distributed and the motion persists
indefinitely, without interruption. Nevertheless, these assumptions
break down in the regime where $z\ll1$. In this limit, the perturbative
term 
\begin{equation}
\frac{\bar{\zeta}}{V_{n}^{z-1}}\sin(\phi_{n+1})\approx\bar{\zeta}V_{n}\left[1-z\ln(V_{n})\right]\sin(\phi_{n+1})\label{limZless1}
\end{equation}
tends to vanish as $V_{n}\leadsto0$, introducing a nonlinear suppression
of motion at low velocities. This suppression is particularly significant
when $\phi_{n}\approx0\,\mod\pi$, where $\sin(\phi_{n+1})\approx0$,
further inhibiting the system's ability to sustain motion. As a result,
the system may become trapped in a dynamically frozen state. This
behavior signals the occurrence of a dynamical phase transition into
a frozen regime when the adiabatic parameter falls below a critical
threshold $z_{c}(\bar{\zeta})$, even though naive scaling arguments
would predict an enhancement of superdiffusion for $z<1$. 

Under the condition that $\phi_{n}\approx0\,(\mathrm{mod}\,\pi)$,
the trapping condition can thus be expressed by examining the magnitude
of the velocity increment in the map equations \eqref{Smap-1-1}:
\begin{equation}
\left|V_{n+1}-V_{n}\right|=\left|\Delta V\right|\approx V_{n}^{1-z}\bar{\zeta}\eta,
\end{equation}
where we have defined $\eta=\sin(\phi_{n+1})\ll1$. This relation
indicates that, under these circumstances, the perturbation is insufficient
to significantly alter the velocity, causing the orbit to remain trapped
near $V_{n}$.

To estimate the minimum perturbation $\Delta V_{\text{min}}$ necessary
for escape from the trapping region, we require: $\Delta V_{\text{min}}\gtrsim V_{n},$leading
to the condition 
\begin{equation}
\bar{\zeta}\eta\gtrsim V_{n}^{z}.
\end{equation}
Solving for $z$, we obtain the critical trapping threshold: 
\begin{equation}
z\lesssim z_{c}(\bar{\zeta})\approx\frac{\ln(\eta\bar{\zeta})}{\ln(V_{n})}\approx\frac{\ln(\eta)}{\ln(V_{n})}+\frac{\ln(\bar{\zeta})}{\ln(V_{n})}.\label{CriticalZ}
\end{equation}
where it is taken into account that, from a numerical standpoint,
both $\eta$ and $V$ are small, on the order of $10^{-3}$, and can
be adjusted in accordance with numerical simulations. 

This modified bouncer map is formally a twist map, because $\frac{\partial\phi_{n+1}}{\partial V_{n}}=2$
for all $(\phi_{n},V_{n})$, but it dynamically behaves as a non-twist
map in the low-velocity (freezing) regime. The effective rotation
number, $\Omega(V_{n})=2\langle V_{n}\rangle,$ approaches zero in
the freezing phase $(z<z_{c}(\bar{\zeta}))$, becoming flat across
initial conditions and leading to $\frac{d\Omega}{dV}\to0$. This
behavior characterizes non-twist maps, known for supporting shearless
barriers, reconnection zones, and enhanced stickiness.

\begin{figure}
\centering{}\includegraphics[width=0.95\columnwidth]{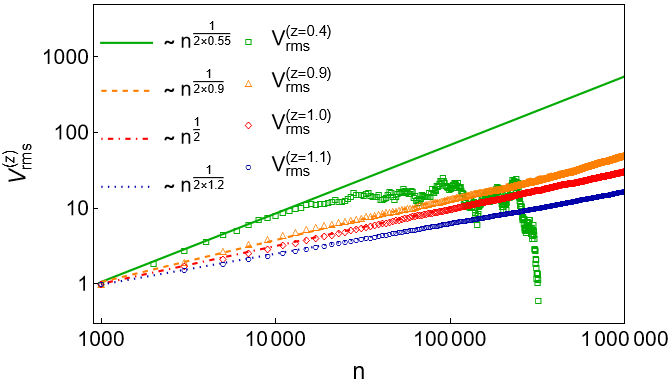}\caption{Plots of $V_{\text{rms}}$ for four values of $z$ are shown, based
on an ensemble of $10^{4}$ initial conditions uniformly distributed
in $\phi\in[0,2\pi]$ and evolved over $10^{6}$ iterations. To enable
meaningful comparison across diffusive regimes, all curves are normalized
such that $V_{\text{rms}}^{(z)}(1000)=1$. Simulations use $\bar{\zeta}=5$
with $z=1.1$, $1.0$, $0.9$, and $0.429$. Solid (green), dashed
(orange), dot-dashed (red), and dotted (blue) lines represent the
theoretical predictions from Eq.(\ref{V2-2}).}
\label{FigVrms}
\end{figure}

\textit{Numerical Analysis} - To substantiate the theoretically predicted
dynamical behaviors, extensive numerical simulations were performed,
consisting of $n=10^{6}$ iterations for each of $10^{4}$ distinct
initial conditions. The choice of $n=10^{6}$ ensures that the long-term
behavior of the system is accurately captured, while the use of $10^{4}$
distinct initial conditions mitigates the influence of specific trajectories
The observable under consideration is the root-mean-square velocity,
computed over an ensemble of $M$ initial conditions and defined as
$V_{\text{rms}}^{(z)}(n)=\sqrt{\left\langle V_{n}^{2}\right\rangle }=\sqrt{\frac{1}{M}\sum_{i=1}^{M}V_{i,n}^{2}}$.
These simulations were conducted for four representative values of
the $z$ parameter: namely, $z=1.1$, $z=1.0$, $z=0.9$, and $z=0.4$,
with $\bar{\zeta}=5$, where the invariant curves have already been
eliminated \citep{Leonel2021}. According to Eq. (\ref{V2-2}), the
expected asymptotic behaviors for the observable are as follows: for
$z=1.1$, the growth follows $\sim n^{1/2.2}$; for $z=1.0$, $\sim n^{1/2}$;
for $z=0.9$, $\sim n^{1/1.8}$; and for $z=0.4$, $\sim n^{1/0.8}$.

Figure~\ref{FigVrms} displays the numerical results corresponding
to the cases $z=1.1$, $z=1.0$, and $z=0.9$, which demonstrate good
agreement with the theoretical predictions. The observed behavior
corroborates the expected scaling laws, thereby validating the analytical
framework presented. For $z=0.4$, the root-mean-square velocity initially
displays superdiffusive growth, approximately following the predicted
power law. However, beyond $n\gtrsim10^{5}$, $V_{\text{rms}}^{(z)}$
undergoes a rapid decline towards zero, indicating the onset of a
dissipative mechanism that drives the system into a dynamically frozen
state. This transition to a frozen state is analogous to a sublimation
process.

Eq.~(\ref{CriticalZ}) predicts a transition to a dynamically frozen
state when the adiabatic $z$ parameter drops below a critical threshold.
This is confirmed numerically for $z=0.4$ and $\bar{\zeta}=5$, as
shown in figure \ref{FigVrms}. To further investigate this transition,
we construct a ``phase'' diagram that elucidates the dependence
of the system's dynamical behavior on the parameters $\bar{\zeta}$
and $z$.

This analysis involves generating a two-dimensional grid of discrete
parameter values, with $\bar{\zeta}$ ranging from 0.2 to 5.0 and
$z$ spanning the interval from 0.2 to 1.2. For each pair $(\bar{\zeta},z)$
in this grid, we compute the corresponding root-mean-square velocity
$V_{\text{rms}}^{(z)}$. Eq. (\ref{V2-2}) shows that $V_{\text{rms}}$
evolves as a power-law function of the number of iterations $n$,
i.e., $V_{\text{rms}}=an^{b},$ where $a$ is a constant and $b$
is the diffusion exponent characterizing the system's long-term behavior.

The exponent $b$ serves as a diagnostic for classifying the diffusion
regime. Specifically, for $b\leq0.49$, the system exhibits subdiffusive
behavior, characterized by dispersion slower than that of normal diffusion..
In the range $0.49<b<0.51$, the diffusion is approximately normal,
implying a linear growth of mean-square velocity with time. Finally,
for $b\geq0.51$, the system enters the superdiffusive regime, where
the transport process is faster than in the normal diffusive case.

\begin{figure}
\centering{}\includegraphics[width=0.9\columnwidth]{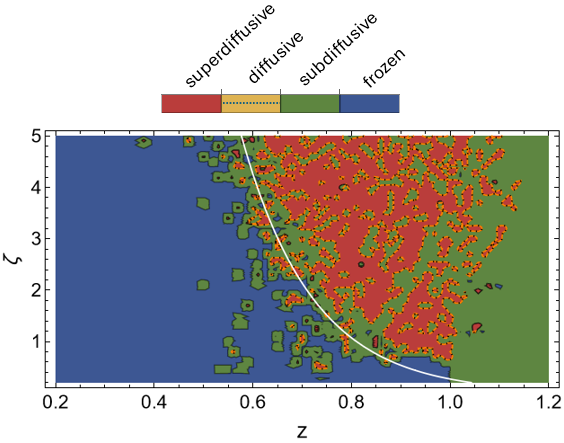}\caption{The phase diagram is categorized by the power-law relation $V_{\text{rms}}=an^{b}$.
Subdiffusion occurs for $b\protect\leq0.49$, superdiffusion for $b\protect\geq0.51$,
and normal diffusion is identified in the intermediate range $0.49<b<0.51$,
marked by a dotted boundary. The white line indicates the theoretical
prediction from Eq. (\ref{CriticalZ}) with $\eta=0.0037$ and $V_{n}=0.001$.
\label{FigDifusiv}}
\end{figure}

These classifications and their corresponding behaviors are illustrated
in the accompanying figure \ref{FigDifusiv}, which visually represents
the varying diffusion regimes as a function of the parameters $\zeta$
and $z$. 
Given that normal diffusive behavior is confined to a narrow range
of $b$ values, it is represented by dashed lines at the boundaries
between subdiffusive (green region) and superdiffusive regimes (red
region). The blue region represents the frozen phase. The white curve
in the figure corresponds to the theoretical relationship 
\begin{equation}
z_{c}=0.86-0.217\ln(\bar{\zeta}),
\end{equation}
as given by Eq.\textasciitilde (\ref{CriticalZ}). This expression
was obtained using the parameter values $\eta=0.019$ and $V_{n}=0.01$,
which are consistent with the assumption that both $\eta$ and $V_{n}$
are small. 

Naturally, this theoretical result\textemdash derived from a simplified
dimensional analysis\textemdash does not capture the detailed structure
of the boundary. Nevertheless, it provides a remarkably good approximation
of the overall trend observed in the boundary as revealed by numerical
simulations showed in figure \ref{FigDifusiv}. It should be emphasized
that once the system enters the frozen regime, increasing the number
of iterations does not restore diffusive behavior. Likewise, for fixed
control parameters $\bar{\zeta}$ and $z$, increasing the number
of initial conditions\textemdash even beyond $10^{4}$\textemdash does
not alter the macroscopic dynamical state. Both the diffusive and
frozen phases thus exhibit robustness with respect to ensemble size
and integration time.

The transition boundaries observed numerically are neither smooth
nor sharply defined, reflecting the nonlinear nature of the system,
which includes fixed points and stability islands embedded in a chaotic
phase space. These structures lead to abrupt dynamical changes under
small variations in $z$ and $\bar{\zeta}$, emphasizing the system\textquoteright s
sensitivity to the parameters. To quantify the geometric complexity
of the transition boundary, the Box-Counting Method \citep{Peitgen2004}
was employed. The frontier dimension $D$ is estimated via the scaling
relation: $D=\lim_{\epsilon\to0}\frac{\log N(\epsilon)}{\log(1/\epsilon)},$
where $\epsilon$ is the box size and $N(\epsilon)$ is the number
of boxes required to cover the boundary. The interface delimiting
the normal diffusive regime exhibits a dimension of $D\approx1.517$,
revealing a nontrivial fractal geometry. This structure is consistent
with critical behavior near dynamical phase transitions and reflects
scale-invariant organization in phase space \citep{Grebogi1983,Tel2008,Zaslavsky2002}.
These manifestations of criticality and scale invariance serve to
characterize the underlying dynamical transition to the frozen phase.

\begin{figure}[t]
\begin{centering}
\includegraphics[width=0.9\columnwidth]{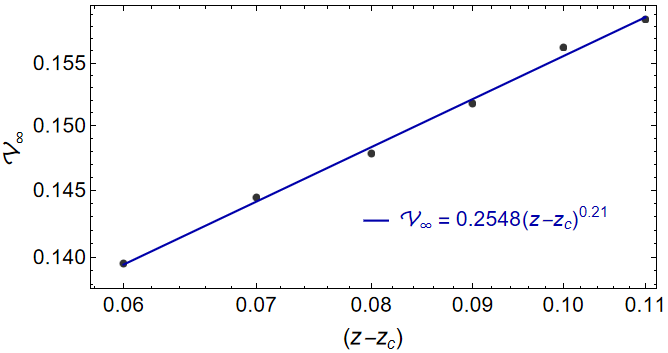}
\par\end{centering}
\caption{Log-Log Plot of $\mathcal{V}_{\infty}$ by $(z-z_{c})$, with $\bar{\zeta}=5$
and $z_{c}=0.55$.}
\label{FigVinf}
\end{figure}

\textit{Dynamical Phase Transition} - This adiabatic bouncer map exhibits
a rich variety of diffusive behaviors depending on the parameters
$\bar{\zeta}$ and $z$, especially a dynamical phase transition to
a frozen state as the $z$ parameter decreases below the critical
value $z_{c}(\bar{\zeta})$. This transition is characterized by the
vanishing of the long-term energy growth, leading to a dynamically
trapped or ``frozen'' regime. To quantify this transition, we define
the \textit{order parameter} as the asymptotic average velocity per
iteration: 
\begin{equation}
\mathcal{V}_{\infty}=\lim_{n\to\infty}\ll V_{n}\gg,
\end{equation}
where $\ll V_{n}\gg=\sqrt{\frac{\left\langle V_{n}^{2}\right\rangle }{n}}.$
This quantity is positive for $z>z_{c}$ and vanishes continuously
as $z\to z_{c}^{+}$. The critical behavior of $\mathcal{V}_{\infty}$
near the transition is captured by the scaling law: 
\begin{equation}
\mathcal{V}_{\infty}\sim(z-z_{c})^{\beta},\qquad\text{for }z\to z_{c}^{+},\label{Vinf}
\end{equation}
where $\beta$ is the critical exponent associated with the phase
transition. This is analogous to magnetization $M$ in ferromagnetic
systems. The numerical results for $\mathcal{V}_{\infty}$ as a function
of $(z-z_{c})$ with $\bar{\zeta}=5$ and $z_{c}=0.55$ are presented
in Figure \ref{FigVinf}, providing support for the theoretical prediction
given by Eq.(\ref{Vinf}). 

The dynamical susceptibility is then defined as
\begin{equation}
\chi(z)\sim\frac{d}{dz}(z-z_{c})^{\beta}=\beta(z-z_{c})^{\beta-1}.\label{sucept}
\end{equation}
The numerical analysis yields a critical exponent $\beta=0.21$, indicating
a divergence of the dynamical susceptibility, as described by Eq.
(\ref{sucept}), in the limit $z\to z_{c}^{+}$. This behavior is
characteristic of a second-order (continuous) phase transition.

\textit{Conclusions and outlook -} This study demonstrates that the
adiabatic parameter plays a critical role in governing both the energy
exchange mechanisms and the underlying dynamical structure of the
system. Variations in this parameter give rise to distinct dynamical
regimes\textemdash namely, subdiffusion, normal diffusion, and superdiffusion\textemdash and
culminate in a second-order phase transition from diffusive behavior
to a frozen state. The presence of fractal boundaries separating these
diffusive regimes highlights a pronounced sensitivity to initial conditions
and parameter values, underscoring the intricate and complex organization
of the system's phase space.

The twist-to-non-twist transition marks a topological change linked
to the breakdown of KAM tori, enriching the system\textquoteright s
dynamical behavior. These findings have broader relevance to systems
driven by adiabatic interactions and nonlinear transport.

As a perspective for future research, the adiabatic framework developed
herein may be directly extended to dynamical billiard systems with
time-dependent boundaries. In particular, a connection with thermodynamic
concepts\textemdash through a formal definition of entropy\textemdash offers
a promising avenue for exploration, with potential implications for
one- and two-dimensional gas models. Furthermore, generalizations
incorporating stochastic perturbations, inter-particle interactions,
or external driving forces could provide deeper insights into the
robustness and universality of the dynamical phase transitions identified
in this work.

\textit{Acknowledgments}\textemdash The author thanks Prof. Edson
Denis Leonel for his valuable contributions to this work.

\bibliographystyle{apsrev}
\addcontentsline{toc}{section}{\refname}\bibliography{references}

\begin{thebibliography}{18}
\expandafter\ifx\csname natexlab\endcsname\relax\def\natexlab#1{#1}\fi
\expandafter\ifx\csname bibnamefont\endcsname\relax
  \def\bibnamefont#1{#1}\fi
\expandafter\ifx\csname bibfnamefont\endcsname\relax
  \def\bibfnamefont#1{#1}\fi
\expandafter\ifx\csname citenamefont\endcsname\relax
  \def\citenamefont#1{#1}\fi
\expandafter\ifx\csname url\endcsname\relax
  \def\url#1{\texttt{#1}}\fi
\expandafter\ifx\csname urlprefix\endcsname\relax\def\urlprefix{URL }\fi
\providecommand{\bibinfo}[2]{#2}
\providecommand{\eprint}[2][]{\url{#2}}

\bibitem[{\citenamefont{Strogatz}(2018)}]{Strogatz2018}
\bibinfo{author}{\bibfnamefont{S.~H.} \bibnamefont{Strogatz}},
  \emph{\bibinfo{title}{Nonlinear Dynamics and Chaos: With Applications to
  Physics, Biology, Chemistry, and Engineering}} (\bibinfo{publisher}{CRC
  Press}, \bibinfo{year}{2018}).

\bibitem[{\citenamefont{Ott}(2002)}]{Ott2002}
\bibinfo{author}{\bibfnamefont{E.}~\bibnamefont{Ott}},
  \emph{\bibinfo{title}{Chaos in Dynamical Systems}}
  (\bibinfo{publisher}{Cambridge University Press}, \bibinfo{year}{2002}).

\bibitem[{\citenamefont{Klages et~al.}(2008)\citenamefont{Klages, Radons, and
  Sokolov}}]{Klages2008}
\bibinfo{editor}{\bibfnamefont{R.}~\bibnamefont{Klages}},
  \bibinfo{editor}{\bibfnamefont{G.}~\bibnamefont{Radons}}, \bibnamefont{and}
  \bibinfo{editor}{\bibfnamefont{I.~M.} \bibnamefont{Sokolov}}, eds.,
  \emph{\bibinfo{title}{Anomalous Transport: Foundations and Applications}}
  (\bibinfo{publisher}{Wiley-VCH}, \bibinfo{year}{2008}).

\bibitem[{\citenamefont{Metzler and Klafter}(2000)}]{Metzler2000}
\bibinfo{author}{\bibfnamefont{R.}~\bibnamefont{Metzler}} \bibnamefont{and}
  \bibinfo{author}{\bibfnamefont{J.}~\bibnamefont{Klafter}},
  \bibinfo{journal}{Physics Reports} \textbf{\bibinfo{volume}{339}},
  \bibinfo{pages}{1} (\bibinfo{year}{2000}).

\bibitem[{\citenamefont{Venegeroles}(2009)}]{Venegeroles2009}
\bibinfo{author}{\bibfnamefont{R.}~\bibnamefont{Venegeroles}},
  \bibinfo{journal}{Physical Review Letters} \textbf{\bibinfo{volume}{102}},
  \bibinfo{pages}{064101} (\bibinfo{year}{2009}).

\bibitem[{\citenamefont{Zaburdaev et~al.}(2015)\citenamefont{Zaburdaev,
  Denisov, and Klafter}}]{Zaburdaev2015}
\bibinfo{author}{\bibfnamefont{V.}~\bibnamefont{Zaburdaev}},
  \bibinfo{author}{\bibfnamefont{S.}~\bibnamefont{Denisov}}, \bibnamefont{and}
  \bibinfo{author}{\bibfnamefont{J.}~\bibnamefont{Klafter}},
  \bibinfo{journal}{Reviews of Modern Physics} \textbf{\bibinfo{volume}{87}},
  \bibinfo{pages}{483} (\bibinfo{year}{2015}).

\bibitem[{\citenamefont{Liebchen et~al.}(2011)\citenamefont{Liebchen, Diakonos,
  and Schmelcher}}]{Liebchen2011}
\bibinfo{author}{\bibfnamefont{B.}~\bibnamefont{Liebchen}},
  \bibinfo{author}{\bibfnamefont{F.~K.} \bibnamefont{Diakonos}},
  \bibnamefont{and}
  \bibinfo{author}{\bibfnamefont{P.}~\bibnamefont{Schmelcher}},
  \bibinfo{journal}{New Journal of Physics} \textbf{\bibinfo{volume}{13}},
  \bibinfo{pages}{093039} (\bibinfo{year}{2011}).

\bibitem[{\citenamefont{Leonel and McClintock}(2004)}]{Leonel2004}
\bibinfo{author}{\bibfnamefont{E.~D.} \bibnamefont{Leonel}} \bibnamefont{and}
  \bibinfo{author}{\bibfnamefont{P.~V.~E.} \bibnamefont{McClintock}},
  \bibinfo{journal}{Journal of Physics A: Mathematical and General}
  \textbf{\bibinfo{volume}{37}}, \bibinfo{pages}{8937} (\bibinfo{year}{2004}).

\bibitem[{\citenamefont{Chirikov}(1979)}]{Chirikov1979}
\bibinfo{author}{\bibfnamefont{B.~V.} \bibnamefont{Chirikov}},
  \bibinfo{journal}{Physics Reports} \textbf{\bibinfo{volume}{52}},
  \bibinfo{pages}{263} (\bibinfo{year}{1979}).

\bibitem[{\citenamefont{Zaslavsky}(2002)}]{Zaslavsky2002}
\bibinfo{author}{\bibfnamefont{G.~M.} \bibnamefont{Zaslavsky}},
  \bibinfo{journal}{Physics Reports} \textbf{\bibinfo{volume}{371}},
  \bibinfo{pages}{461} (\bibinfo{year}{2002}).

\bibitem[{\citenamefont{Huang}(1987)}]{Huang1987}
\bibinfo{author}{\bibfnamefont{K.}~\bibnamefont{Huang}},
  \emph{\bibinfo{title}{Statistical Mechanics, 2nd Edition}}
  (\bibinfo{year}{1987}),
  \urlprefix\url{https://ui.adsabs.harvard.edu/abs/1987stme.book.....H}.

\bibitem[{\citenamefont{Leonel and Livorati}(2015)}]{Leonel2015}
\bibinfo{author}{\bibfnamefont{E.~D.} \bibnamefont{Leonel}} \bibnamefont{and}
  \bibinfo{author}{\bibfnamefont{A.~L.~P.} \bibnamefont{Livorati}},
  \bibinfo{journal}{Communications in Nonlinear Science and Numerical
  Simulations} \textbf{\bibinfo{volume}{20}}, \bibinfo{pages}{159}
  (\bibinfo{year}{2015}), ISSN \bibinfo{issn}{1007-5704}, \eprint{1405.2030},
  \urlprefix\url{https://ui.adsabs.harvard.edu/abs/2015CNSNS..20..159L}.

\bibitem[{\citenamefont{Cecconi et~al.}(2022)\citenamefont{Cecconi, Costantini,
  Taloni, and Vulpiani}}]{Cecconi2022}
\bibinfo{author}{\bibfnamefont{F.}~\bibnamefont{Cecconi}},
  \bibinfo{author}{\bibfnamefont{G.}~\bibnamefont{Costantini}},
  \bibinfo{author}{\bibfnamefont{A.}~\bibnamefont{Taloni}}, \bibnamefont{and}
  \bibinfo{author}{\bibfnamefont{A.}~\bibnamefont{Vulpiani}},
  \bibinfo{journal}{Physical Review Research} \textbf{\bibinfo{volume}{4}},
  \bibinfo{eid}{023192} (\bibinfo{year}{2022}), \eprint{2206.06786},
  \urlprefix\url{https://ui.adsabs.harvard.edu/abs/2022PhRvR...4b3192C}.

\bibitem[{\citenamefont{Boscolo et~al.}(2023)\citenamefont{Boscolo, Junior, and
  Barreiro}}]{Boscolo2023}
\bibinfo{author}{\bibfnamefont{A.~L.} \bibnamefont{Boscolo}},
  \bibinfo{author}{\bibfnamefont{V.~B. d.~S.} \bibnamefont{Junior}},
  \bibnamefont{and} \bibinfo{author}{\bibfnamefont{L.~A.}
  \bibnamefont{Barreiro}}, \bibinfo{journal}{Physical Review E}
  \textbf{\bibinfo{volume}{107}}, \bibinfo{pages}{045001}
  (\bibinfo{year}{2023}), ISSN \bibinfo{issn}{2470-0053}.

\bibitem[{\citenamefont{Miranda et~al.}(2022)\citenamefont{Miranda, Kuwana,
  Huggler, and et~al.}}]{Leonel2021}
\bibinfo{author}{\bibfnamefont{L.}~\bibnamefont{Miranda}},
  \bibinfo{author}{\bibfnamefont{C.}~\bibnamefont{Kuwana}},
  \bibinfo{author}{\bibfnamefont{Y.}~\bibnamefont{Huggler}}, \bibnamefont{and}
  \bibinfo{author}{\bibnamefont{et~al.}}, \bibinfo{journal}{Eur. Phys. J. Spec.
  Top.} \textbf{\bibinfo{volume}{231}}, \bibinfo{pages}{167}
  (\bibinfo{year}{2022}), ISSN \bibinfo{issn}{1951-6401}.

\bibitem[{\citenamefont{Peitgen et~al.}(2004)\citenamefont{Peitgen,
  J{\"{u}}rgens, and Saupe}}]{Peitgen2004}
\bibinfo{author}{\bibfnamefont{H.}~\bibnamefont{Peitgen}},
  \bibinfo{author}{\bibfnamefont{H.}~\bibnamefont{J{\"{u}}rgens}},
  \bibnamefont{and} \bibinfo{author}{\bibfnamefont{D.}~\bibnamefont{Saupe}},
  \emph{\bibinfo{title}{Chaos and fractals - new frontiers of science {(2.}
  ed.)}} (\bibinfo{publisher}{Springer}, \bibinfo{year}{2004}), ISBN
  \bibinfo{isbn}{978-0-387-20229-7}.

\bibitem[{\citenamefont{Grebogi et~al.}(1983)\citenamefont{Grebogi, Ott, and
  Yorke}}]{Grebogi1983}
\bibinfo{author}{\bibfnamefont{C.}~\bibnamefont{Grebogi}},
  \bibinfo{author}{\bibfnamefont{E.}~\bibnamefont{Ott}}, \bibnamefont{and}
  \bibinfo{author}{\bibfnamefont{J.~A.} \bibnamefont{Yorke}},
  \bibinfo{journal}{Physical Review Letters} \textbf{\bibinfo{volume}{50}},
  \bibinfo{pages}{935} (\bibinfo{year}{1983}), ISSN \bibinfo{issn}{0031-9007}.

\bibitem[{\citenamefont{T{\'e}l and Lai}(2008)}]{Tel2008}
\bibinfo{author}{\bibfnamefont{T.}~\bibnamefont{T{\'e}l}} \bibnamefont{and}
  \bibinfo{author}{\bibfnamefont{Y.-C.} \bibnamefont{Lai}},
  \bibinfo{journal}{Physics Reports} \textbf{\bibinfo{volume}{460}},
  \bibinfo{pages}{245} (\bibinfo{year}{2008}), ISSN \bibinfo{issn}{0370-1573}.

\end{thebibliography}

\end{document}